\documentclass[linenumbers,twocolumn,showpacs,preprintnumbers,amsmath,amssymb,prb,aps]{revtex4}
\usepackage{graphicx}
\usepackage{dcolumn}
\usepackage{bm}
\usepackage{color}
\usepackage{amsmath}

\newif\ifgraph

\graphtrue             %

\begin{document}
\title{Autonomous Ratcheting by Stochastic Resetting}

\author{Pulak K. Ghosh$^{1}$\footnote{pulak.chem@presiuniv.ac.in (corresponding author)}, Shubhadip Nayak$^{1}$, Jianli Liu$^{2}$, Yunyun Li$^{2}$, and Fabio Marchesoni$^{2,3}$\footnote{fabio.marchesoni@pg.infn.it(corresponding author)}}

\affiliation{$^{1}$ Department of Chemistry, Presidency University, Kolkata
700073, India}
\affiliation{$^{2}$ Center for Phononics and Thermal Energy Science, Shanghai
 Key Laboratory of Special Artificial Microstructure Materials and Technology,
 School of Physics Science and Engineering, Tongji University, Shanghai 200092, China}
\affiliation{$^{3}$ Dipartimento di Fisica, Universit\`{a} di Camerino, I-62032 Camerino, Italy}
\date{\today}

\begin{abstract}
We propose a generalization of the stochastic resetting mechanism for a
Brownian particle diffusing in a one-dimensional periodic potential: randomly
in time, the particle gets reset at the bottom of the potential well it was in.
Numerical simulations show that in mirror asymmetric potentials, stochastic
resetting rectifies the particle's dynamics, with
maximum drift speed for an optimal average resetting time. Accordingly, an unbiased
Brownian tracer diffusing on an asymmetric substrate can rectify its motion by
adopting an adaptive stop-and-go strategy. Our proposed ratchet mechanism can model 
directed autonomous motion of molecular motors and micro-organisms 
\end{abstract}
\maketitle

The notion of stochastic resetting (SR) is attracting growing attention (see
Ref. \cite{SR_rev} for a recent review). This term refers to the sudden
interruption of a stochastic process after random time intervals, followed by its
starting anew, possibly after a further latency time, with same initial conditions.
Diffusion under SR is a non-equilibrium stationary process, which found
applications in search contexts \cite{Maj_PRL1}, optimization of randomized
computer algorithms \cite{Zecchina}, and in many biophysical problems
\cite{Reuveni1,Reuveni2}. Surprisingly, under SR the otherwise infinite mean
first passage time of a freely diffusing Brownian
particle \cite{Redner} from an injection point to an assigned target point becomes finite,
and, most notably, can be minimized for an optimal choice of the resetting
time, $\tau$ \cite{Maj_PRL2,Reuveni3}. Many analytical methods earlier
developed in the theory of homogeneous stochastic processes \cite{Redner,Gardiner} can
be generalized to study diffusion under SR, for instance, to calculate the
mean-first-exit time (MFET) of a reset particle out of a one-dimensional (1D)
domain \cite{Pal} or potential well \cite{Reuveni4}. In general, SR speeds
up (slows down) diffusive processes characterized by random escape times
with standard deviation larger (smaller) than the respective averages \cite{Reuveni2}.

\begin{figure}[tp]
\centering \includegraphics[width=8.0cm]{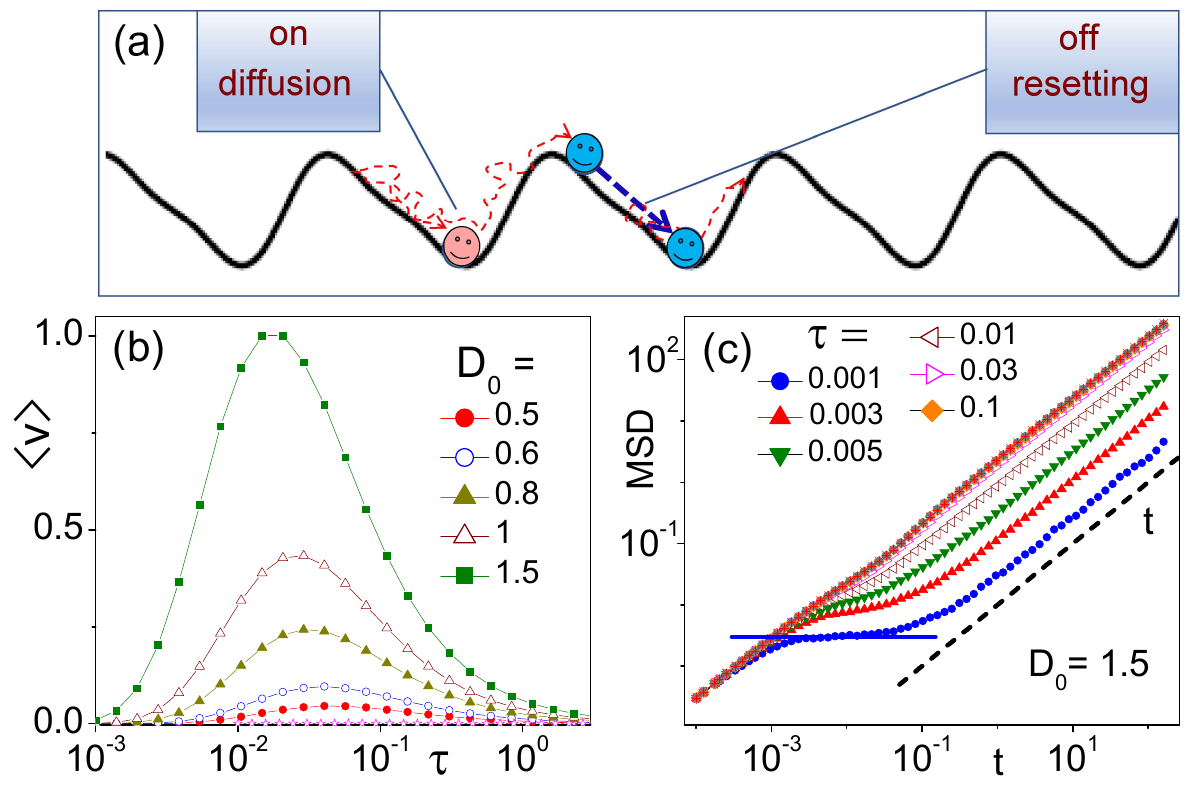}
\caption{Autonomous ratcheting by stochastic resetting: (a) schematics; (b) $\langle v\rangle$
vs $\tau$ for different $D_0$; (c) MSD vs $t$ for $D_0=1.5$ and
different $\tau$. The asymptotic dependence is linear in $t$ (dashed line), while the horizontal
plateaus for very low $\tau$ come close to the square half-width of the relevant
probability density peak, $\langle (x-x_0)^2 \rangle \simeq 2D_0\tau$, given in the text
(see, e.g., the solid line for $\tau=10^{-3}$).
Values of the diffusion constant, $D$, obtained by fitting
Eq. (\ref{fitD}) are plotted in Fig. \ref{F2}(c). Numerical simulations
for the ratchet potential of Eq. (\ref{V}) with $L=1$ and $\tau_0 = 0$. \label{F1}}
\end{figure}
\begin{figure*}[tp]
\centering \includegraphics[width=17.0cm]{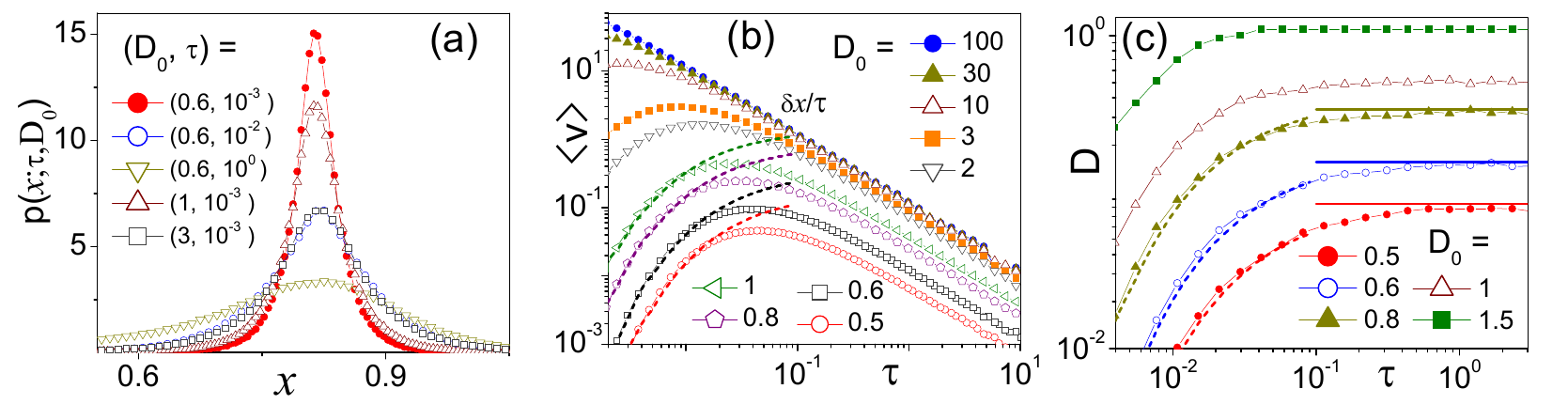} \caption{Simulation
data analysis: (a) $p(x;\tau,D_0)$ for different values of $D_0$ and $\tau$;
(b) $\langle v \rangle$ vs. $\tau$ for different $D_0$. The
dashed curves are our predictions respectively for small $\tau$, $\langle
v\rangle=L/\langle T(\tau) \rangle$, and in the strong noise regime, $\langle
v\rangle=\delta x/\tau$ with $\delta x=(L_L-L_R)/2$; (c) fitting
parameter, $D$, of the diffusion law of Eq. (\ref{fitD}), also for
different $D_0$. Our analytical estimates for
small and large $\tau$ (see text) are represented respectively by dashed and solid curves.
Numerical simulations for the ratchet potential of
Eq. (\ref{V}) with $L=1$ and $\tau_0 = 0$. \label{F2}}
\end{figure*}

In this Letter we propose an SR mechanism with degenerate resetting point.
Let us consider an overdamped Brownian particle of coordinate $x$, diffusing
in a 1D periodic potential, $V(x)$, of period $L$. We assume for simplicity
that the potential unit cells have one minimum each at $x_n=x_0 +nL$, with
$n=0, \pm 1, \dots$. Upon resetting, the particle stops diffusing and falls
instantaneously at the bottom of the potential well it was in; it will resume
diffusing after a latency time $\tau_0 \geq 0$, see Fig. \ref{F1}(a). By this
mechanism of {\em autonomous} SR, we intend to model the dynamics of small
motile tracers (like bacteria or micro-robots \cite{Wang}) capable of
switching their internal engine on and off. In the case of undirected
motility, the tracer would perform an unbiased Brownian motion. Let us further assume that the
barriers separating two adjacent potential minima are asymmetric
under mirror reflection, i.e., $V(x-x_0) \neq V(-x+x_0)$ (ratchet potential
\cite{RMP}). Extensive numerical simulations show that (i) SR {\em rectifies
diffusion} in a ratchet potential. The particle's net drift speed, $\langle v
\rangle$, reaches a maximum for an optimal value of the resetting time, $\tau$,
which strongly depends on the potential profile, see Fig. \ref{F1}(b);
(ii) SR suppresses spatial diffusion. For large observation times, the particle's
mean-square displacement (MSD) turns proportional to time (normal diffusion);
the relevant diffusion constant increases sharply with the resetting time
in correspondence with the maximum of the drift speed, see Fig. \ref{F1}(c).

While this variant of the SR mechanism may be reminiscent of a flashing
ratchet with pulsated temperature \cite{pla}, here the diffusing tracer
exploits the substrate spatial asymmetry to autonomously rectify its
random motion in the absence of external time-dependent fields of force or
gradients \cite{RMP,JCP1,JCP2}, simply by time-operating its internal engine to adjust
to the substrate itself.

\begin{figure*}[tp]
\centering \includegraphics[width=17.5cm]{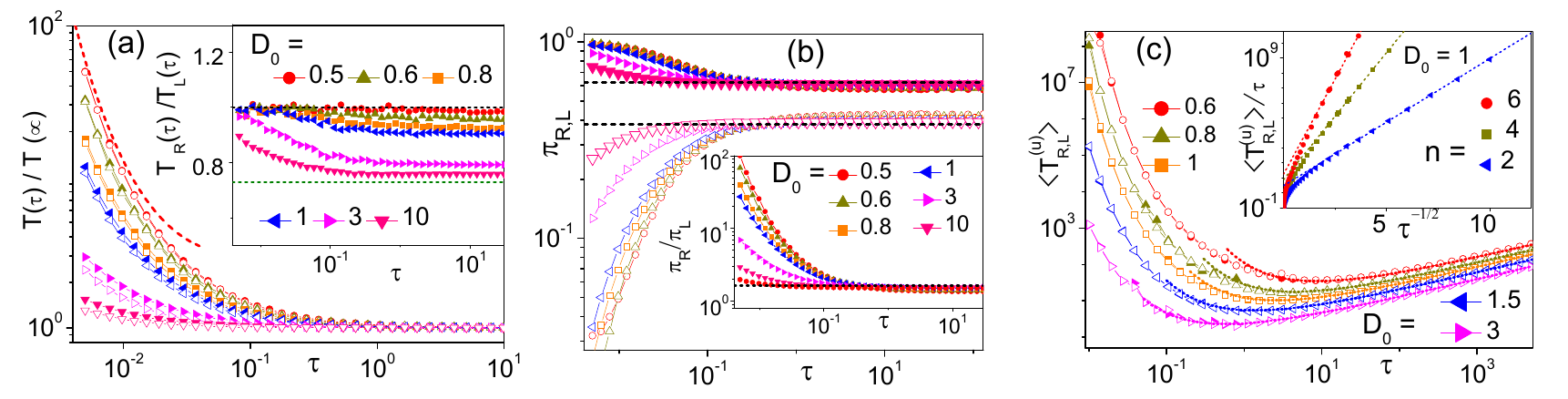}
\caption{Exit time statistics: (a) MFET's out of a potential well through the
right/left barrier, $\langle T_{R,L} \rangle$, or through either barrier, $\langle T
\rangle$; (b) splitting probabilities, $\pi_{R,L}$, for right/left exits (solid/empty symbols), vs
$\tau$ for different $D_0$ (see legends). Note that $\langle T\rangle =\pi_R
\langle T_R \rangle +\pi_L \langle T_L \rangle$ \cite {Redner}. The dashed
curve in (a) represents the analytical estimate of $\langle T(\tau)\rangle
\simeq \langle T_R(\tau)\rangle$ for $\tau\to 0$, Eq. (\ref{T0}),
with $\langle T(\infty) \rangle = T_K$ (see text). The dashed lines in (b)
are the large $D_0$ limits of $\pi_{R,L} (\infty)=L_{L,R}$
(with the corresponding ratio displayed in the inset). The horizontal lines in the inset of (a)
are the expected $\langle T_R \rangle / \langle T_L\rangle$ ratios for $D_0\to 0$ (upper)
and $\infty$ (lower, see text). (c) unconstrained right/left MFPT's, $\langle T^{\rm (u)}_{R,L}(\tau)\rangle$,
$x_0\to x_0 \pm L$, (empty/solid symbols) for different values of $D_0$; inset: small-$\tau$
dependence of the MFPT's for multiple cell transitions, $x_0\to x_0 +nL$, for $D_0=1$. Simulation data are
compared with the relevant estimates of Eqs. (\ref{l_tau}) (large $\tau$) and (\ref{s_tau}) (small $\tau$).
Numerical simulations for the ratchet potential of Eq. (\ref{V}) with $L=1$ and $\tau_0 = 0$. \label{F3}}
\end{figure*}

{\em Model.} The simulated particle dynamics was formulated in terms of the Langevin equation (LE),
\begin{equation}\label{LE}
  \dot x=-V'(x) +\xi(t),
\end{equation}
where $\xi(t)$ denotes a stationary zero-mean valued Gaussian noise with
autocorrelation $\langle \xi(t) \xi(0) \rangle =2D_0\delta(t)$ (white noise)
and $V(x)$ is the standard ratchet potential \cite{RMP},
\begin{equation}\label{V}
  V(x)= \sin(2\pi x/L) +(1/4)\sin(4\pi x/L),
\end{equation}
with asymmetric barriers of height $\Delta V=
(3/2)(1+2/\sqrt{3})^{1/2}\simeq 2.20$. The potential unit cell $[0,L]$ has a
maximum (barrier) at $x_b=(L/2\pi)\arccos[(\sqrt{3}-1)/2]\simeq 0.19L$ and a
minimum (well bottom) at $x_0=L-x_b\simeq 0.81L$, with curvatures
$\omega_0^2=V''(x_0)=-V''(x_b)= (2\pi/L)^2(3\sqrt{3}/2)^{1/2}\simeq
63.6/L^2$, see Fig. \ref{F1}(a). The asymmetric potential wells have
right/left slopes of different lengths,
$L_{R,L}$, with $L_L=x_0-x_b=L-L_R\simeq 0.62L$. In addition to the thermal fluctuations and the ratchet potential, the particle is subjected to resetting to the attracting
 local substrate minimum after a random time that is taken from exponential distribution with mean $\tau = 1/r$.  Where, $r$ is 
the resetting rate. Partly motivated by technical issues in experiments, 
earlier works\cite{Majumdar-A,Majumdar-B,Majumdar-C} 
had considered the case when the particle was reset to a fully randomly chosen position.  
Along with the restart protocol, the Eq.(\ref{LE})  was  numerically
integrated by means of a standard Milstein scheme \cite{Kloeden}, to compute
the drift speed $\langle v\rangle =
\lim_{t\to \infty}[\langle x(t) \rangle -x(0)]/t$, and the asymptotic MSD,
\begin{equation}\label{fitD}
 \langle \Delta x^2(t)\rangle=\langle x^2(t)\rangle -\langle v\rangle^2 t^2 \equiv 2Dt,
\end{equation}
of a particle under stationary conditions (with or without SR), (Figs. \ref{F1} and
\ref{F2}), and the MFET's, $\langle T_{R,L}(\tau)\rangle$, for a reset particle injected
at the bottom of the well, $x_0$, to first exit it through the left (right)
barrier, $x_b$ ($x_b+L$) (Fig. \ref{F3}).

{\em Rectification under SR.} The key features of the resulting SR ratchet are illustrated in the bottom panels of
Fig. \ref{F1}: the particle motion gets rectified with net speed $\langle
v(\tau)\rangle$ [Fig. \ref{F1}(b)] and asymptotic diffusion constant, $D(\tau)$, a
monotonically increasing function of the SR time [Fig. \ref{F1}(c)]. Rectification is maximum in
an optimal $\tau$ range, as $D$ approaches an stationary value (the same as in the absence of SR).

To explain ratcheting  under SR we anticipate two properties of the
statistics of particle escape out of a potential well, summarized in Fig. \ref{F3}. In
the absence of resetting, i.e., for asymptotically large $\tau$, the
probability current density of the process is zero, which rules out rectification
\cite{RMP}, $\langle v(\infty)\rangle=0$. Things change upon decreasing the
SR time, as proven by the $\tau$-dependence of the splitting probabilities,
$\pi_{R,L}(\tau)$, for the particle to exit a potential well through the right/left
barrier. Upon lowering $\tau$, the asymmetry ratio $\pi_R/\pi_L$ in Fig. \ref{F3}(b)
grows monotonically, the effect being more apparent at low noise,
$D_0\ll \Delta V$, so that
$\langle v(\tau)\rangle>0$. We qualitatively explain this property with the
increased asymmetry of the probability density \cite{Redner} of the reset particle around the
potential minima [Fig. \ref{F2}(a)]. On the other hand, the data of Fig. \ref{F3}(b) clearly show that in
the limit $\tau \to 0$, $\langle T_{R}(\tau)\rangle$ diverges exponentially,
so that we anticipate $\langle v(\tau \to 0)\rangle=0+$. The combination of these two opposite effects
determines the typical resonant profile of the
$\langle v(\tau)\rangle$ curves.

{\em Slow SR.} More in detail, the data of Fig. \ref{F2}(b) suggest that $\langle v(\tau)\rangle$
 decays asymptotically like $\tau^{-1}$. This behavior can be easily explained in the strong
noise regime with $D_0 \gg \Delta V$ and $\langle T_{R,L}(\tau)\rangle \ll
\tau$. Under this condition, the particle executes many barrier crossings before being reset at the
bottom of a $V(x)$ well. At resetting, it is caught in average to the left of the
well bottom; hence, at each resetting the particle jumps to the right an average distance,
$\delta x=\bar x -x_0 >0$, $\bar x$ being the center of mass of the
(periodic) particle's stationary probability density function, $p(x; \tau, D_0)$, in the
potential well with bottom at $x=x_0$. Accordingly, the particle gets rectified with positive net drift speed
$\langle v(\tau) \rangle=\delta x/\tau$. In the strong noise regime, $p(x; \tau, D_0)$,
approaches a uniform distribution; hence, $\delta x=(L_L-L_R)/2$,
in good agreement with the numerical data of Fig. \ref{F2}(a).

Upon decreasing the noise strength, $\delta x$ diminishes for two reasons,
as illustrated in Fig. \ref{F2}(a). Firstly, in the absence of SR, i.e., for $\tau \to \infty$, the probability
density, $p(x; \tau, D_0)$, approaches its thermal equilibrium form, $p(x;
\infty, D_0)={\cal N} \exp(-V(x)/D_0)$, with ${\cal N}$ an appropriate
normalization constant. For $D_0 \ll \Delta V$, $p(x; \infty, D_0)$ shrinks
around $x_0$, that is, $\delta x$ diminishes. Secondly, by lowering $D_0$ in the
presence of SR, i.e., for finite $\tau$, $\langle T(\tau)\rangle$ grows
comparable with $\tau$. Accordingly, barrier escape and resetting events grow
 correlated, which invalidates the above estimate of the particle's drift speed.
However, numerical data confirm that $\langle v(\tau)\rangle$, though strongly suppressed,
keeps decaying asymptotically like $1/\tau$, even at low noise.

{\em Fast SR.} The plots of $p(x;\tau,D_0)$ for the lowest $\tau$ values in Fig. \ref{F2}(a)
consist of a central peak tapering off with asymmetric slow-decaying tails on both sides. In the limit
$\tau \to 0$, (i) the peak gets sharper and more symmetric, while remaining
centered at the resetting point, $x_0$. Its square half-width can be easily
calculated for $D_0 \ll \Delta V$, by approximating $V(x)\simeq \omega_0^2(x-x_0)^2/2$ and
averaging over the SR time, that is, $\langle (x-x_0)^2\rangle \simeq
2D_0\tau/(1+2\omega_0^2\tau)$; (ii) the tails get thinner
but more asymmetric. This behavior is consistent with the $\tau$-dependence
of the escape asymmetry ratio, $\pi_R/\pi_L$, displayed in Fig. \ref{F3}(b) \cite{Redner}.

The sharp decay of $\langle v (\tau) \rangle$ for $\tau \to 0$
proves that fast SR eventually suppresses the
interwell particle diffusion. In such limit, as shown in Fig.
\ref{F3},  the particle tends to jump to the right, with
$\pi_R(\tau) \gg \pi_L(\tau)$ and, therefore, $\langle T(\tau)\rangle \simeq
\langle T_R(\tau)\rangle$ with $\langle T(\tau)\rangle \gg \tau$.
Under these conditions, the
resulting drift speed can be easily estimated under renewal theory
approximation \cite{Cox}, that is $\langle v(\tau \rangle=L/\langle T_R (\tau)\rangle$.

To calculate $\langle T_R(\tau)\rangle$ we had recourse to the analytical
results of Ref. \cite{Reuveni4} for Brownian diffusion under SR in the
presence of a constant bias. We made contact with Eq. (6) there, by
replacing the constant bias with the effective (right-to-left)
restoring force of our ratchet potential, $\Delta V/L_R$. In the limit $\tau
\to 0$, the MFET for the transition to the adjacent well on the right,
$x_0\to x_0+L$, is twice the MFET for the transition $x_0\to x_b+L$, that is
\begin{equation}\label{T0}
   \langle T(\tau)\rangle \simeq \langle T_R(\tau)\rangle \simeq 2\tau \exp\left ({\Delta V}/{2D_0}+{L_R}/{\sqrt{D_0\tau}} \right ).
 \end{equation}
Of course, this approximation holds good only for $\pi_R(\tau) \simeq 1$
($\pi_L(\tau)\simeq 0$), and its agreement with the numerical data improves
upon decreasing the noise strength, i.e., for $D_0 \lesssim \Delta V$, as
shown in Fig. \ref{F3}(a). On making use of this estimate for $\langle
T_R(\tau)\rangle$, we closely reproduced also
the raising branches of the $\langle v(\tau)\rangle$ curves in Fig. \ref{F2}(b).

{\em Diffusion under SR.} Regarding the intrawell diffusion,
we remind that in the absence of SR, the MFET from
$x_0$ to $x_0 \pm L$ amounts to the standard Kramers' time \cite{Gardiner}
$T_K=(2 \pi/\omega_0^2)\exp(-\Delta V/D_0)$. By the same token, one concludes
that for $\tau \to \infty$, $\langle T_R \rangle \simeq \langle T_L \rangle$, with both MFET's
tending to $T_K$ for $D_0/\Delta V \to 0$, and their
ratio, $\langle T_R\rangle /\langle T_L \rangle$
approaching $(1+L/L_L)/(1+L/L_R)\simeq 0.72$ in the opposite limit,
 $D_0/\Delta V\to \infty$.  On the other hand, for large $\tau$ the splitting
probabilities can be easily computed assuming no SR
(see Sec. 5.2.7 of Ref. \cite{Gardiner}); their limits for $D_0/\Delta V\to 0$ (and $\to \infty$)
are respectively $\pi_{R,L}(\infty) =1/2$ (and $L_{L,R}/L$), as shown in Fig. \ref{F3}(b).

These remarks are useful to interpret the MSD data sets of Fig. \ref{F1}(c).
Numerical simulation indicates that diffusion
at large times, $t\gg \langle T(\tau) \rangle$, is normal, as anticipated by the fitting law
of Eq. (\ref{fitD}). At small $\tau$, a transient plateau for
$t \lesssim \langle T(\tau)\rangle$,  $\langle \Delta x^2 \rangle\simeq 2D_0\tau $,
marks the particle relaxation inside a single potential well [with $\langle \Delta x^2 \rangle$
of the order of the square half-width of the $p(x;\tau, D_0)$ peak estimated above]. The $\tau$-dependence of the
asymptotic diffusion constants, $D$, is reported in Fig. \ref{F2}(c). For large $\tau$, the
$D(\tau)$ curves approach the horizonal asymptotes \cite{Gardiner}, $D=L^2/2T_K$,
as to be expected in the absence of SR. Vice versa for very short SR times,
the diffusion constant is well approximated by $D(\tau)=L^2/2\langle T_R(\tau)\rangle$,
as predicted by the renewal theory for a process with average escape time constant $\langle T_R(\tau)\rangle$
\cite{Cox}. In both $\tau$ limits, our phenomenological arguments are supported by
numerical simulation.

\begin{figure}[tp]
\centering \includegraphics[width=8.0cm]{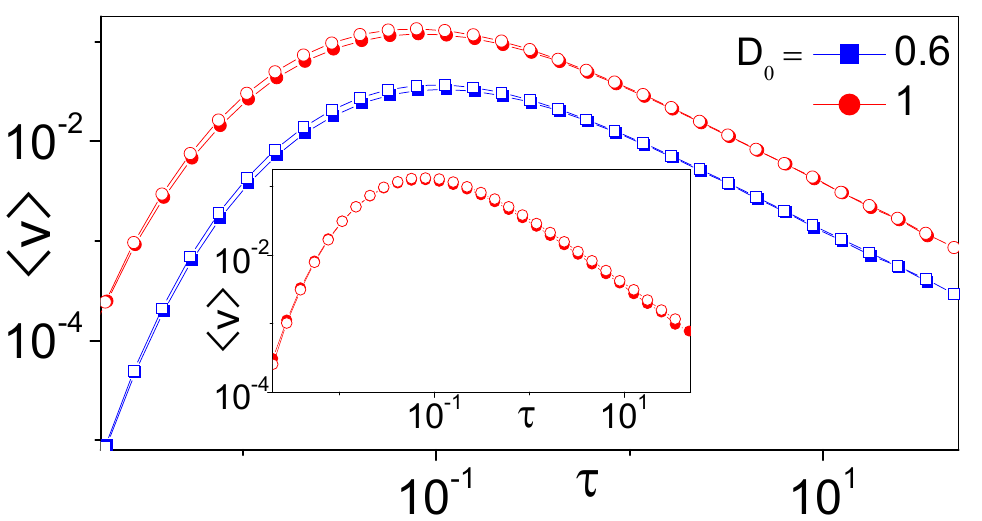}
\caption{Rectification speed, $\langle v(\tau)\rangle$, of the SR ratchet of Fig. \ref{F1} with
$\tau_0=0.1$ (filled symbols) and $\tau_0=0$ (empty symbols), for different $D_0$. The zero-latency data
have been rescaled according to Eq. (\ref{lat}). Inset: $\langle v(\tau) \rangle$ of
a flashing ratchet with dichotomic noise strength, $D_0(t)$, switching between 0
(fixed waiting time $\tau_0=0.1$) and $D_0$ (random waiting times
exponentially distributed with average $\tau$), compared with the flashing ratchet
in the main panel (circles).
\label{F4}}
\end{figure}

{\em Comparison with standard diffusion under SR.}
Numerical data in Fig. \ref{F3}(a) show that by decreasing the SR time, $\langle
T_R(\tau)\rangle$ keeps being larger than $\langle
T_L(\tau)\rangle$. Moreover, $\langle T(\tau)\rangle$ grows
monotonically with $\tau$, i.e., the MFET out of the potential well is not optimized by
resetting. Of course, the predicted SR optimization of the average passage times
\cite{Maj_PRL2} is still detectable, but only for the unconstrained transitions $x_0\to
x_b$ with $x\geq x_b$, and $x_0\to x_b+L$ with $x\leq x_b+L$. In panel (c) of
Fig. \ref{F3} we investigated the same transitions as in panel (a), except
for the reflecting barriers, which were shifted at $\mp \infty$. The
corresponding right/left unconstrained MFPT curves, $\langle T^{\rm
(u)}_{R,L}(\tau)\rangle$, overlap throughout the entire $\tau$ range. Furthermore,
all MFPT curves diverge for $\tau \to \infty$, as to be expected due to the lack
of a reflecting barrier \cite{Maj_PRL2}. In the absence of SR (i.e.,
for $\tau \to \infty$), the particle still diffuses over the substrate like a
free particle, but with the reduced effective diffusion constant, $D=L^2/2T_K$,
of Eq. (\ref{fitD}) [see fits in Fig. \ref{F2}(c)]. This suggests rewriting
Eq. (7) of Ref. \cite{Maj_PRL2} as
\begin{equation}\label{l_tau}
  \langle T^{\rm (u)}_{R,L}(\tau)\rangle/\tau=\exp{(L/\sqrt{D\tau})}-1,
\end{equation}
a formula that well reproduces the large-$\tau$ branches of the curves in Fig
\ref{F3}(c) with no additional fitting parameters. In the inset of the same figure, we
analyze the small-$\tau$ dependence of the MFPT's for the right transitions $x_0\to
x_0 + nL$ with $n=1,2, \dots$ and reflecting barriers at $-\infty$.
By applying the heuristic argument invoked to derive Eq. (\ref{T0}), we
obtain the working approximate estimate,
\begin{equation}\label{s_tau}
   \langle T^{\rm (u)}_{R,L}(\tau)\rangle/\tau \simeq \exp ({\Delta V}/{2D_0}+{nL}/{\sqrt{D_0\tau}}),
 \end{equation}
which holds for $n$-cell transitions to the right/left at vanishingly small $\tau$. Note that here,
contrary to Eq. (\ref{s_tau}), we make use of the free diffusion constant, $D_0$.

{\em Concluding remarks.} The SR ratcheting mechanism introduced above can be readily generalized to  more realistic
cases when resetting takes a finite time \cite{SR_rev}, $\tau_0$, called
here latency time. The relevant net ratchet speed turns out to be a function of both $\tau$ and
$\tau_0$, $\langle v(\tau, \tau_0)\rangle$, which can be related
with the zero-latency speed, $\langle v(\tau, 0)\rangle$, through a simple time
rescaling, namely
\begin{equation}\label{lat}
  \langle v(\tau, \tau_0)\rangle=\langle v(\tau,0)\rangle/(1+\tau_0/\tau)),
\end{equation}
as illustrated in Fig. \ref{F4}(a).

This instance of SR ratchet lends itself to a simple laboratory demonstration.
We start again from the LE (\ref{LE}) with the potential of Eq. (\ref{V}) but,
instead of implementing the SR protocol with latency
time $\tau_0$, we now assume a dichotomic noise strength, $D_0(t)$, with
$D_0=0$ for fixed time intervals, $\tau_0$, and $D_0(t)=D_0$ for random time
intervals exponentially distributed with average $\tau$. The resulting LE
describes a rectifier, which could be classified as a special case of flashing
ratchet \cite{pla}. In one regard the two rectification mechanisms are
apparently similar: in both cases the particle rests at the bottom of a potential well
for the time interval, $\tau_0$, before resuming Brownian diffusion,
because either reset that way (SR ratchet) or given enough time to relax
there (flashing ratchet with $\omega_0^2 \tau_0 \gg 1$). As shown in the
inset of Fig. \ref{F4}, for the same choice of the tunable parameters, $D_0,
\tau$ and $\tau_0$, the rectification power of the two ratchets is almost
identical. Therefore, one can utilize a ratchet with dichotomic noise strength
to experimentally demonstrate the rectification properties of the proposed SR ratchet.
However, an important difference between these two ratchets is also noteworthy.
The flashing ratchet is fueled by an external source capable of ``heating and
cooling'' the particle or its substrate \cite{Libchaber,Sano}. SR ratcheting
with finite latency time,
instead, can be controlled by the particle itself, by autonomously regulating
its own internal motility mechanism for maximum efficiency.

In summary, we have proposed a new protocol of stochastic resetting,
whereby a particle diffusing on a one-dimensional substrate, gets reset not
at a fixed point, but rather at one of the degenerate minima of the
substrate. We investigated, both numerically and analytically, the diffusion
properties of the reset particle and showed that for spatially asymmetric
substrates the particle gets rectified with direction determined by the
substrate profile, and optimal speed depending on the resetting time. We
argue that, thanks to such a mechanism, a motile system (biological and synthetic, alike)
can exploit the substrate asymmetry to autonomously direct its motion,
for instance, by
randomly switching on and off its propulsion engine at an appropriate rate.

\section*{Acknowledgements}
Y.L. is supported by the NSF China under grants No. 11875201 and No.
11935010. P.K.G. is supported by SERB Core Research Grant No. CRG/2021/007394.
\section*{Data Availability}
The data that support the findings of this study are available within the article.
\section*{Conflict of interest}
The authors have no conflicts to disclose.

\end{document}